\documentclass[twocolumn,aip,jcp,reprint,floatfix,amsmath,amssymb]{revtex4-1}
\usepackage{natbib}
\usepackage{amsmath,amsthm}
\usepackage{enumerate}
\usepackage{graphicx}
\usepackage{subfigure}
\usepackage{tabularx}
\usepackage{threeparttable}
\usepackage{dcolumn}
\usepackage{bm}
\usepackage{color,epsfig,multirow}
\usepackage{array}
\usepackage{hyperref}
\usepackage{algorithm}
\usepackage{algorithmic}
\usepackage{upgreek}

\newtheorem{theorem}{Theorem}[section]

\renewcommand{\epsilon}{\varepsilon}
\theoremstyle{definition}

\theoremstyle{remark}

\allowdisplaybreaks[4]

\begin{document}
\preprint{Preprint}

\title{Superscalability of the random batch Ewald method}


\author{Jiuyang Liang}
\altaffiliation{These authors contributed equally.}
\affiliation{School of Mathematical Sciences, Shanghai Jiao Tong University, Shanghai 200240, China}

\author{Pan Tan}
\altaffiliation{These authors contributed equally.}
\affiliation{School of Physics and Astronomy and Institute of Natural Sciences, Shanghai Jiao Tong University, Shanghai 200240, P. R. China}

\author{Yue Zhao}
\affiliation{School of Mathematical Sciences, Shanghai Jiao Tong University, Shanghai 200240, China}

\author{Lei Li}
\affiliation{School of Mathematical Sciences, Institute of Natural Sciences and MoE-LSC,
	Shanghai Jiao Tong University, Shanghai 200240, China}

\author{Shi Jin}
\affiliation{School of Mathematical Sciences, Institute of Natural Sciences and MoE-LSC,
	Shanghai Jiao Tong University, Shanghai 200240, China}

\author{Liang Hong}
\altaffiliation{Corresponding author.\\ Electronic address: hongl3liang@sjtu.edu.cn, xuzl@sjtu.edu.cn.}
\affiliation{School of Physics and Astronomy and Institute of Natural Sciences, Shanghai Jiao Tong University, Shanghai 200240, P. R. China}

\author{Zhenli Xu}
\altaffiliation{Corresponding author.\\ Electronic address: hongl3liang@sjtu.edu.cn, xuzl@sjtu.edu.cn.}
\affiliation{School of Mathematical Sciences, Institute of Natural Sciences and MoE-LSC,
Shanghai Jiao Tong University, Shanghai 200240, China}

\renewcommand{\thefootnote}{\fnsymbol{footnote}}

\footnotetext[1]{Corresponding author}

\begin{abstract}
Coulomb interaction, following an inverse-square force-law, quantifies the amount of force between two stationary and electrically charged particles. The long-range nature of Coulomb interactions poses a major challenge to molecular dynamics simulations which are major tools for problems at the nano-/micro- scale. Various algorithms are developed to calculate the pairwise Coulomb interactions to a linear scaling but the poor scalability limits the size of simulated systems. Here, we conduct an efficient molecular dynamics algorithm with the random batch Ewald method on all-atom systems where the complete Fourier components in the Coulomb interaction are replaced by randomly selected mini-batches. By simulating the $N$-body systems up to 100 million particles using $10$ thousand CPU cores, we show that this algorithm furnishes $O(N)$ complexity, almost perfect scalability and an order of magnitude faster computational speed when compared to the existing state-of-the-art algorithms. Further examinations of our algorithm on distinct systems, including pure water, micro-phase-separated electrolyte and protein solution demonstrate that the spatiotemporal information on all time and length scales investigated and thermodynamic quantities derived from our algorithm are in perfect agreement with those obtained from the existing algorithms. Therefore, our algorithm provides a breakthrough solution on scalability of computing the Coulomb interaction. It is particularly useful and cost-effective to simulate ultra-large systems, which was either impossible or very costing to conduct using existing algorithms, thus would benefit the broad community of sciences.
\end{abstract}


\maketitle
%
\section{Introduction}
Molecular dynamics (MD) is one of the most powerful simulation tools in modern science to furnish the atomic-detailed microscopic mechanism underlying experimental findings in a plethora of areas including physics, chemistry, engineering, biology and pharmaceutical sciences \cite{RN3,RN1,RN9,RN10,RN8}. Despite the enormous success, the application of MD simulation without specific coarse graining and enhanced sampling method has been largely limited to moderate size (often below 1 million atoms) and time scale (shorter than 10 microseconds). These limitations cannot be solved by parallel computing using a large number of computational cores, as the inter-atomic Coulomb interactions are long-ranged and require intensive communications between cores, significantly reducing the parallel efficiency, especially when using supercomputers \cite{RN10,RN68,RN70}. In the past decades, enormous efforts have been devoted to reduce the computational cost of the Coulomb interaction, and many important algorithms, including the lattice summation methods on the basis of fast Fourier transform (FFT) \cite{RN13,RN14} and multipole type methods such as the tree code \cite{RN11} and the fast multipole method (FMM)  \cite{RN12}, have been developed. Both FFT and FMM were named among the top 10 algorithms in scientific computing developed in the 20th century\cite{RN37}, which can reduce the computational complexity to $O(N\mathrm{log}N)$ or even $O(N)$, and have achieved a great success and been widely applied in the main-stream MD packages. However, none of these methods can achieve high scalability for calculating of the Coulomb interaction when using a large number of computational cores\cite{RN18}.

The random batch Ewald method (RBE)\cite{RN23} is an alternative and promising $O(N)$ algorithm for electrostatic calculations.  It is based on the Ewald splitting, but it avoids the use of the FFT, instead random mini-batch sampling on the Fourier space is introduced to approximate the force contribution from the long-range part. In this work, we develop the RBE for the all-atom molecular dynamics simulation and demonstrate its great computational efficiency and high scalability, especially so when applied on large-scale simulations using supercomputer owing to great reduction in global communications. The so-called ``random mini-batch'' used in the RBE, originated from the stochastic gradient descent method widely used in machine learning \cite{RN71,RN22}, was first proposed for interacting particle systems with rigorous error estimates\cite{RN24}, and has succeeded in Monte Carlo simulation on particle systems\cite{RN40}. In the present work, we demonstrate that the RBE-based MD enhances the computational speed by an order of magnitude in comparison to the state-of-the-art algorithms including particle-particle particle-mesh (PPPM) and particle-mesh Ewald (PME) methods, and maintains the parallel efficiency of nearly $95\%$ when paralleling up to $10,000$ computational cores to simulate a large system of $100$ million atoms. Moreover, a systematic test was conducted on all-atom simulations of three representative systems: bulk water, micro-phase separated aqueous electrolyte and protein solution, and it reveals that the spatiotemporal information for these systems on all time and length scales and the thermodynamical quantities derived from the PPPM and PME are quantitatively reproduced by the RBE-based MD. Thus, as compared to the mainstream algorithms (PPPM or PME\cite{RN13,RN14}), which are widely used in MD simulation of all kinds of molecular systems, the RBE-based MD furnishes a novel algorithm with substantial improvement in computational efficiency and parallel scalability while maintaining the same accuracy of the spatiotemporal information.

This paper is organized as follows: In Section \ref{methods}, we introduce the random batch Ewald method, and then describe the parallel implementation in details. In Section \ref{results}, we validate the superior CPU performance and superscalability of the RBE method by the simulation on bulk water, and perform numerical calculations for three benchmark problems, including bulk water, micro-phase separated aqueous electrolyte, and protein solution, to demonstrate the accuracy of the RBE method. Discussions are given in Section \ref{discussions}.

\section{Methods}\label{methods}
\subsection{The random-batch Ewald method}
In the classical Ewald method\cite{RN25}, the Coulomb kernel is split into two components,
\begin{equation}
	\frac{1}{r}=\frac{\operatorname{erf}(\sqrt{\alpha} r)}{r}+\frac{\operatorname{erfc}(\sqrt{\alpha} r)}{r}
\end{equation}
where $\text{erf}{\left(\cdot\right)}$ is the error function and $\text{erfc}(\cdot)$ is its complementary function, such that the first term is a smooth function and the second one becomes short-ranged. To mimic the bulk environment, a periodic boundary condition is assumed. Without loss of generality, one considers a system of $N$ charged particles located at $\bm{r}_j$ for $j=1,\cdots,N$ in a cubic box of length $L$ and the volume of the box is given by $V=L^3$. The Coulomb force acting on the $i$-th particle in the form of the Ewald summation is,
\begin{equation}
	\begin{aligned}
		\boldsymbol{F}_{i}=&- \sum_{\boldsymbol{k} \neq 0} \frac{4 \pi q_{i} \boldsymbol{k}}{V |\bm{k}|^{2}} e^{-\frac{|\bm{k}|^{2}}{4 \alpha}} \operatorname{lm}\left(e^{-i \boldsymbol{k} \cdot \boldsymbol{r}_{i}} \rho(\boldsymbol{k})\right)\\
		&-q_{i} \sum_{j, \bm{n}}{ }^{\prime} q_{j} G\left(\left|\boldsymbol{r}_{i j}+\boldsymbol{n} L\right|\right) \frac{\boldsymbol{r}_{i j}+\boldsymbol{n} L}{\left|\boldsymbol{r}_{i j}+\boldsymbol{n} L\right|} \\
		=:& \boldsymbol{F}_{i, 1}+\boldsymbol{F}_{i, 2}
	\end{aligned}
\end{equation}
where $\bm{r}_{ij}=\bm{r}_j-\bm{r}_i$ is the vector starting at particle $i$ and pointing towards particle $j$, $\bm{k}=2\pi \bm{m}/L$ with $\bm{m}\in\mathbb{Z}^{3}$, $G(r)$ and $\rho(\bm{k})$ are functions defined by
\begin{equation}\label{eq3}
	G(r):=\frac{\operatorname{erfc}(\sqrt{\alpha} r)}{r^{2}}+\frac{2 \sqrt{\alpha} \mathrm{e}^{-\alpha r^{2}}}{r^{2}}, \quad \rho(\boldsymbol{k}):=\sum_{i=1}^{N} q_{i} e^{i \boldsymbol{k} \cdot \bm{r}_{i}}.
\end{equation}
Here the structure factor $\rho\left(\bm{k}\right)$ is the conjugate of the Fourier transform of the charge density. In Eq.~\eqref{eq3}, $\bm{F}_{i,1}$ represents the component in the Fourier space while $\bm{F}_{i,2}$ denotes the one in real space. By proper choice of the parameters ($\alpha$ and the real-space and Fourier-space cutoffs $r_c$ and $k_c$), the computational complexity is optimized to $O(N^{3/2})$. Moreover, the FFT is often employed to further speed up the evaluation of $\bm{F}_{i,1}$ such that the cutoff radius can be much smaller, resulting in the core algorithms for mainstream software, including the PPPM and PME algorithms\cite{RN13,RN14}. A final computational complexity of $O(N\log N)$ can be achieved through these methods for periodic systems.

The RBE avoids the use of the FFT, instead it employs the random mini-batch strategy to calculate $\bm{F}_{i,1}$. Here, one picks a small batch of frequencies when evaluating $\bm{F}_{i,1}$ for a given particle $i$. These frequencies are chosen randomly with an importance-sampling scheme (see below for a brief overview and for more details in Ref. \cite{RN23}). Let $P$ be the batch size, $\bm{k}_\ell$ be the $\ell$-th frequency and
\begin{equation}
	S:=\sum_{\bm{k}\neq0}{\exp(-|\bm{k}|^2/4\alpha)},
\end{equation}
then the approximate force reads,
\begin{equation}\label{force}
	\boldsymbol{F}_{i, 1}^{*}=-\sum_{\ell=1}^{P} \frac{S}{P} \frac{4 \pi \boldsymbol{k}_{\ell} q_{i}}{V |\bm{k}_{\ell}|{ }^{2}} \operatorname{lm}\left(e^{-i \boldsymbol{k}_{\ell} \cdot \boldsymbol{r}_{i}} \rho\left(\boldsymbol{k}_{\ell}\right)\right).
\end{equation}

\begin{figure*}[htbp]	
	\centering
	\includegraphics[width=1\textwidth]{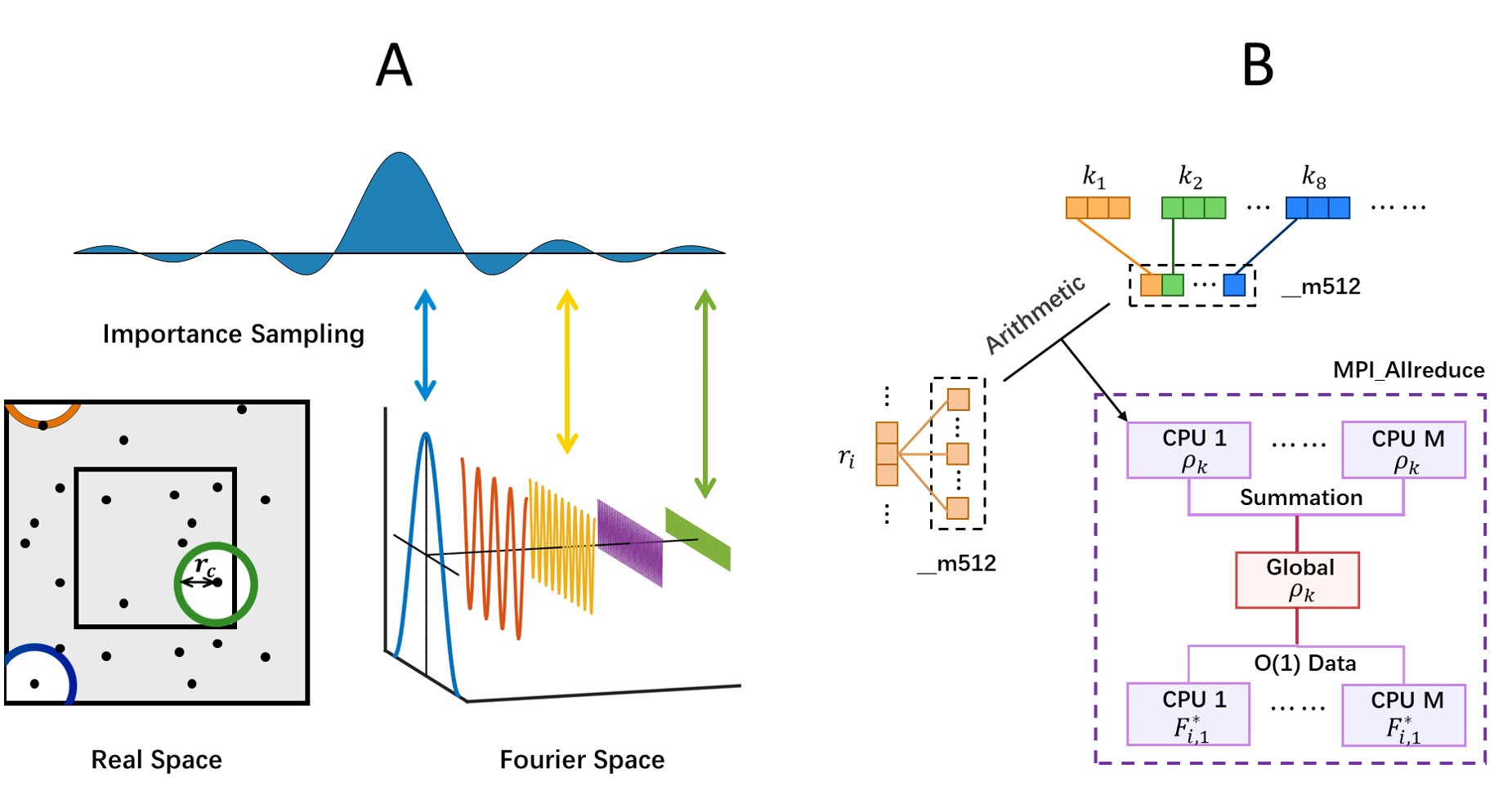}
	\caption{\textbf{Sketch map}. (A): The algorithm sketch map of the RBE applied to electrostatic long-range interaction. (B): Parallel strategy in the Fourier space employing SIMD. Only one global operation with $O(1)$ amount of data is required during each step. }
	\label{fig:1}
\end{figure*}

Since a system-size independent $P$ number of frequencies is used to estimate $\rho\left(\bm{k}\right)$ for each particle, the overall computational cost to calculate the structure factor for the entire simulation system scales as $O(N)$. The sketch map of the RBE algorithm is present in Fig.~\ref{fig:1}A.

In the isothermal-isobaric ensemble simulations (NPT), the virial tensor calculation is needed at each MD step. It can also be done by employing importance sampling in the Fourier space. With the same batch of the $P$ Fourier modes, the approximate reciprocal space contribution of the virial reads,
\begin{equation}\label{pressure}
	\begin{split}
	\Xi_{1}^{\beta \gamma}=-\sum_{\ell=1}^{P} \frac{S}{P} \frac{\pi}{V |\bm{k}_{\ell}|^{2}}&\left|\rho\left(\boldsymbol{k}_{\ell}\right)\right|^{2}\cdot\\
	&\left[\delta_{\beta \gamma}-2 k_{\ell}^{\beta} k_{\ell}^{\gamma}\left(\frac{1}{4 \alpha^{2}}+\frac{1}{|\bm{k}_{\ell}|^{2}}\right)\right]
	\end{split}
\end{equation}
where $\beta$ and $\gamma$ are dimensions taken from the three coordinates indicating the corresponding components of the tensor. The details for deriving Eq.~\eqref{pressure} are briefly given in Section~\ref{virial}. The steps of the RBE-based MD algorithm are summarized as follows:
\begin{enumerate}[(i)]
	\item Set parameters $\alpha$, $r_c$ and $k_c$, and batch size $P$. Load initial positions and strengths of charges.
	\item Sample sufficient number of $\bm{k}\sim e^{-|\bm{k}|^2/4\alpha}$, $\bm{k}\neq\bm{0}$ with $\bm{0}$ the zero vector, by employing importance sampling via the MH procedure to form the total set of frequency samples, $\mathcal{K}$.
	\item Evolve the Newton's equations. The real part $\bm{F}_{i,2}$ of the Coulomb force is directly computed with cutoff $r_c$, whereas the Fourier part is approximated by $\bm{F}_{i,1}^\ast$ using Eq.~\eqref{force} with the $P$ frequencies chosen from $\mathcal{K}$ in order.
	\item If the NPT ensemble is employed, compute the real-space virial and the approximated Fourier virial using Eq.~\eqref{pressure}.
\end{enumerate}

\subsection{Parallel implementation}
The RBE is especially suitable for parallelization and vectorization. Here we present the implementation strategy with hybrid MPI/OpenMP parallelization for the RBE in both the all-atom NVT and NPT simulations, which supports massively parallel MD simulations of large-scale systems. We use the Intel 512-bit SIMD (AVX-512 architecture) for vectorization implementation, which operates sixteen neighbors for single-precision floating calculation (or eight for double precision) at the same time, and the Intel Parallel Studio for parallelization (including functions of MPI, OpenMP, and AVX512 instructions). The communication and vectorization procedures are optimized as follows.

Step (ii) and Step (iii) of the RBE require a serial importance sampling procedure and a global broadcast operation, whereas their cost is relatively small and can be eliminated by the designed non-jammed communication and computation/communication overlapping. First, for the NVT ensemble, assume that $M$ MPI ranks are employed, and $M$ independent sampling processes are executed in parallel within each rank. Next, the first MPI rank broadcasts the samples to other ranks using blocking operation. Finally, the computation step (iii) of the summarized RBE-based MD algorithm in the main paper is executed, whereas the samples in other ranks is concurrently broadcasted. This strategy evaluates and updates the samples every $M$ steps, dramatically reducing the global communication cost. Second, for the NPT ensemble, the above strategy is no longer works due to the dynamically changed size of the box. We offer an alternative method by the lights of the Multiple-Program Multiple-Data parallelization in GROMACS \cite{RN15}. When $M$ is large, one MPI rank is selected to do only the sampling which is sampled from the standard normal distribution and then broadcasts the samples and the random variables which are required in the Metropolis step to other ranks. Other ranks receive the samples and multiply by a constant with respect to the instantaneous size of the box on them. An acceptance-rejection step is then run at each rank with the same random variables. The communication operation can also overlap with the computation.

Step (iii) also requires the evaluation of the real-part force $\bm{F}_{i,2}$ of the Coulomb force and we follows the classical procedure in the MD package of LAMMPS\cite{RN17,RN36}. The calculation of the Fourier part ($\bm{F}_{i,1}^\ast$) is rather time-consuming in previous methods, and it is now evaluated using Eq.\eqref{force}, whose parallel strategy is displayed in Fig.~\ref{fig:1}B. The samples and the positions of particles are packaged into 512-bit vectors when the structure factors $\rho\left(\bm{k}\right)$ are evaluated. Only one global operation, MPI$\_$Allreduce, is required for reducing $\rho\left(\bm{k}\right)$. The approximated force $\bm{F}_{i,1}^\ast$ of each particle and the Fourier virial are then obtained from the structure factors.

\subsection{Calculation of the virial}\label{virial}
We briefly discuss the calculation of virial Eq.~\eqref{pressure}. In a system with periodic boundary conditions, the macroscopic pressure $p$ of a set of $N$ particles contained in a volume $V$ has the following well-established relation
\begin{equation}
	p=\frac{2}{3 V}\left(E_{\text{kin}}-\Xi\right)
\end{equation}
where $E_{\text{kin}}$ is the instantaneous kinetic energy and $\Xi$ is the virial. In the full tensor form, the virial can be written as
\begin{equation}\label{inner}
	\Xi=-\frac{1}{2} \sum_{i<j}^{N} \boldsymbol{r}_{i j}^{n} \otimes \bm{F}_{i j},
\end{equation}
where $\bm{r}_{ij}^n$ denotes the distance vector of the nearest image of atom $i$ from atom $j$ and $\bigotimes$ denotes the direct product of two vectors. The straightforward implementation of Eq.~\eqref{inner} involves its evaluation in the inner loop of the non-bonded force routine, which results in a significant CPU time consumption. Nevertheless, it is possible to extract the virial calculation from the inner loop. A hybrid method is often employed to evaluate the Ewald-based electrostatic virial that the real-space virial is treated as the tensor form employing the method in the literature\cite{RN48} and the Fourier virial is computed from the derivative of Lagrangian $U_F$
\begin{equation}\label{realpres}
	\begin{split}
		&\Xi_{1}^{\beta \gamma}=-\sum_{\eta} \frac{-\partial U_{\mathrm{F}}}{\partial h_{\gamma \eta}} h_{\beta \eta}\\
		&=-\sum_{\bm{k} \neq \bm{0}} \frac{\pi}{V |\bm{k}|^{2}} e^{-\frac{|\bm{k}|^{2}}{4 \alpha}}|\rho(\boldsymbol{k})|^{2}\left[\delta_{\beta \gamma}-2 k^{\beta} k^{\gamma}\left(\frac{1}{4 \alpha}+\frac{1}{|\bm{k}|^{2}}\right)\right]
	\end{split}
\end{equation}
where $h$ is the tensor indicating the size of the box, $\beta$, $\gamma$, $\eta$ are taken from $\{x,y,z\}$ indicating the corresponding component. Although the Fourier virial can be cheaply evaluated using Eq.~\eqref{realpres}, it cannot be directly derived from RBE due to the incomplete data of the structure factors, as only $P$ of them is evaluated in Eq.~\eqref{force} in the main paper. To address this problem, we follow the same idea that employs importance sampling from the Gaussian distribution in the Fourier space. With the same batch of $P$ frequencies of $\bm{k}$ (see Eq.~\eqref{force}), the approximate reciprocal space contribution of virial $\Xi_1$ reads as Eq.~\eqref{pressure}.

\subsection{Consistency and stability analysis of employing RBE in NPT ensemble}
In this part, we will give a brief analysis of the consistency and stability of employing RBE in NPT ensemble. More discussions will be reported in our subsequent work.

We define the fluctuation in the random batch approximation for the Fourier part of the virial on particle $i$ by
\begin{equation}
	\chi^{\beta \gamma}=\Xi_{1}^{\beta \gamma}-\widetilde{\Xi}_{1}^{\beta \gamma}.
\end{equation}
The expectation and variance of the fluctuation can be obtained by direct calculation, which is given by
\begin{equation}
	\begin{aligned}
		&\mathbb{E}\left(\chi^{\beta \gamma}\right)=0\\
		&\mathbb{E}\left(\left|\chi^{\beta \gamma}\right|^{2}\right)=\frac{S}{P}\sum_{|\bm{k}| \neq |\bm{0}|} \frac{\pi^{2}|\rho(\bm{k})|^{4}}{V^{2} |\bm{k}|^{4}} e^{-\frac{|\bm{k}|^{2}}{4 \alpha}}\mathcal{Q}_{\beta\gamma}(|\bm{k}|)-\left|\Xi_{1}^{\beta \gamma}\right|^{2}
	\end{aligned}
\end{equation}
where
\begin{equation}
	\mathcal{Q}_{\beta\gamma}(|\bm{k}|)=\delta_{\beta \gamma}-2 k^{\beta} k^{\gamma}\left(\frac{1}{4 \alpha^{2}}+\frac{1}{|\bm{k}|^{2}}\right)^{2}.
\end{equation}

The following result indicates that the RBE is valid for capturing the finite time dynamics of the NPT ensemble (we take the Langevin thermostat\cite{RN31} and the C-rescale barostat\cite{RN47} for illustration).
\begin{theorem}
	Let $(\bm{r}_i,\ \bm{v}_i,\ \varepsilon)$ be the solutions to
	\begin{equation}
		\begin{gathered}
			d \boldsymbol{r}_{i}=\boldsymbol{v}_{i} d t \\
			m_{i} d \boldsymbol{v}_{i}=\left[\boldsymbol{F}_{i}\left(\left\{\boldsymbol{r}_{i}\right\}\right)-\gamma_{T} \boldsymbol{v}_{i}\right] d t+\sqrt{\frac{2 \gamma_{T}}{k_{\text{B}} T}} d \boldsymbol{W}_{i} \\
			d \varepsilon=-\frac{\beta_{T}}{\gamma_{p}}\left(P_{0}-\frac{2}{3 e^{\varepsilon}}\left(E_{k i n}-\Xi\right)\right) d t+\sqrt{\frac{2 k_{\text{B}} T \beta_{T}}{e^{\varepsilon} \gamma_{p}}} d \boldsymbol{W}_{\varepsilon},
		\end{gathered}
	\end{equation}
where $\left\{\bm{W}_i\right\}$ and $\bm{W}_\varepsilon$ are i.i.d. Wiener processes, $k_{\text{B}}$ is the Boltzmann constant, $\varepsilon$ is the strain defined as $\log(V)$, $\beta_T$ is an estimate of the isothermal compressibility of the system, $\gamma_T$ and $\gamma_p$ are two characteristic times associated to the thermostat and the barostat, respectively, and $P_0$ is the external pressure. Let $({\widetilde{\bm{r}}}_i,\ {\widetilde{\bm{v}}}_i,\ \widetilde{\varepsilon})$ be the solutions to
\begin{equation}
	\begin{gathered}
		d \tilde{\boldsymbol{r}}_{i}=\widetilde{\boldsymbol{v}}_{i} d t \\
		m_{i} d \widetilde{\boldsymbol{v}}_{i}=\left[F_{i, 1}^{*}\left(\left\{\tilde{\boldsymbol{r}}_{i}\right\}\right)+F_{i, 2}\left(\left\{\tilde{\boldsymbol{r}}_{i}\right\}\right)-\gamma_{T} \widetilde{\boldsymbol{v}}_{i}\right] d t+\sqrt{\frac{2 \gamma_{T}}{k_{B} T}} d \boldsymbol{W}_{i} \\
		d \tilde{\varepsilon}=-\frac{\beta_{T}}{\gamma_{p}}\left(P_{0}-\frac{2}{3 e^{\tilde{\varepsilon}}}\left(E_{k i n}-\widetilde{\Xi}\right)\right) d t+\sqrt{\frac{2 k_{B} T \beta_{T}}{e^{\tilde{\varepsilon}} \gamma_{p}}} d \boldsymbol{W}_{\varepsilon},
	\end{gathered}
\end{equation}
with the same initial values as $(\bm{r}_i,\bm{v}_i,\varepsilon)$. Suppose that the masses $m_i$ are bounded uniformly from above and below. If the forces $\bm{F}_i$ are bounded and Lipschitz and $\mathbb{E}\left(\chi^{\beta\gamma}\right)=0$, then for any simulation time $\mathcal{T}>0$, there exists $C(\mathcal{T})>0$ such that
\begin{equation}
	\begin{split}
	\mathbb{E}\Bigg{[}\frac{1}{N} \sum_{i}\bigg{(}\left|\boldsymbol{r}_{i}-\tilde{\boldsymbol{r}}_{i}\right|^{2}+\left|\boldsymbol{v}_{i}-\widetilde{\boldsymbol{v}}_{i}\right|^{2}&+|\varepsilon-\tilde{\varepsilon}|^{2}\bigg{)}\Bigg{]}^{\frac{1}{2}} \\
	&\leq C(N, \mathcal{T}) \sqrt{\Lambda \delta t},
	\end{split}
\end{equation}
where
\begin{equation}
	\Lambda=\max \left(E\left(\left|F_{i, 1}(\boldsymbol{r})-F_{i, 1}^{*}(\boldsymbol{r})\right|^{2}\right), \mathbb{E}\left(\left|\chi^{\beta \gamma}\right|^{2}\right)\right)
\end{equation}
is an upper bound of the variance.
\end{theorem}
Similar proofs for interacting particle systems can be found in the literature\cite{RN24,RN40}, and we omit here.

\section{Results}\label{results}
In this section, we perform the main results of this paper to demonstrate that the RBE-based MD enhances the computational speed by an order of magnitude in comparison to the state-of-the-art algorithms including PPPM and PME, and maintains promising parallel efficiency. Moreover, a systematic test was conducted on all-atom simulations of three systems: bulk water, micro-phase separated aqueous electrolyte and protein solution, to demonstrate the accuracy on the spatiotemporal information and the thermodynamical quantities of RBE-based MD. The experimental design for all these simulations and informations of the employed hardware and software are given in Appendices \ref{appendix:experimental} and \ref{appendix:hardware}, respectively.

\subsection{CPU performance}
\begin{figure}[htbp]	
	\centering
	\includegraphics[width=0.5\textwidth]{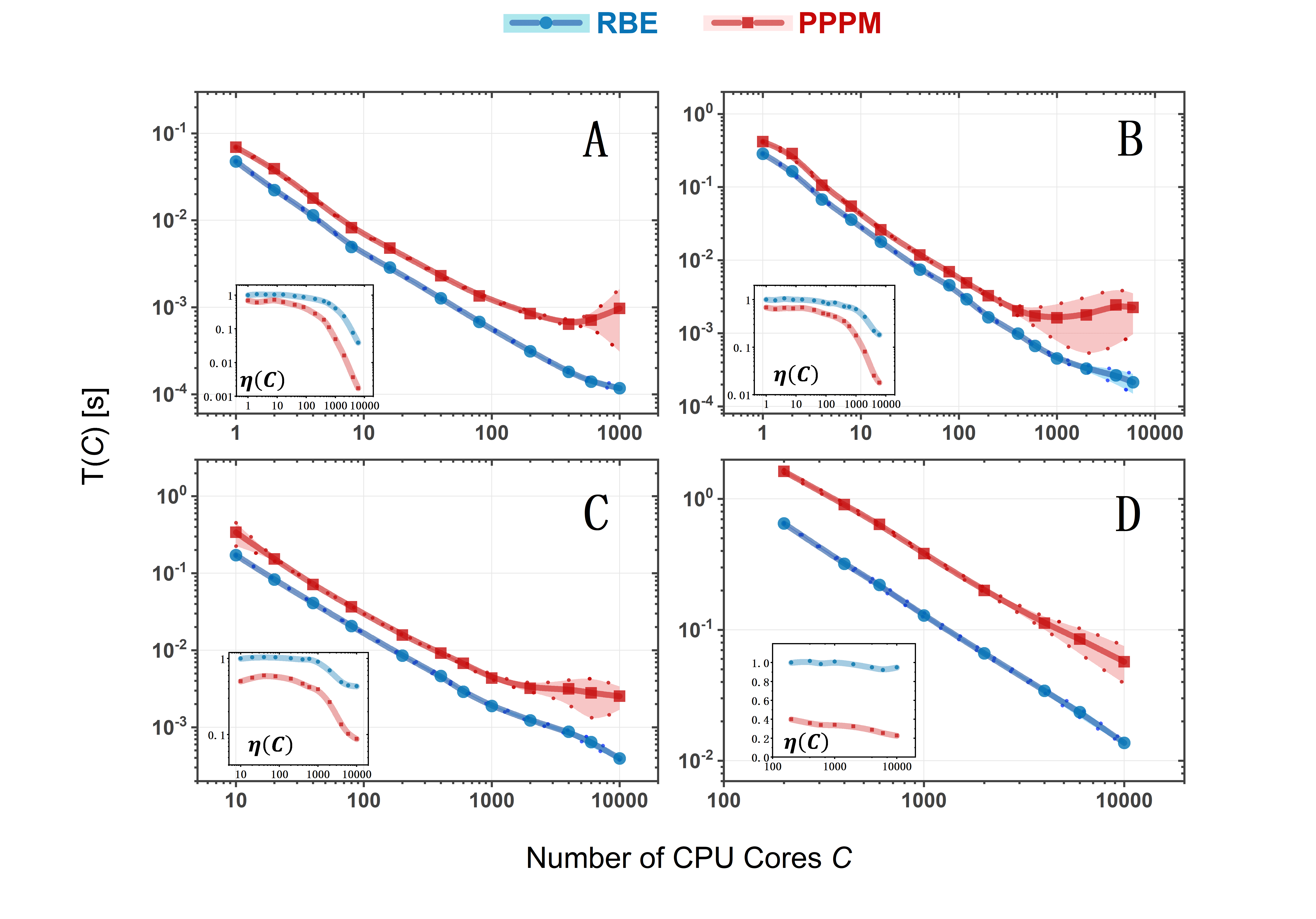}
	\caption{\textbf{Performance Comparison}. $T(C)$, i.e., the average CPU time spent per simulation step for four bulk water systems of different sizes of (A) $53367$ atoms, (B) $311469$ atoms, (C) $3000000$ atoms, and (D) $100158744$ atoms, as a function of the number of CPU cores, for the RBE and the PPPM. Blue line: RBE with $P=100$. Red line: PPPM with $\Delta={10}^{-4}$. The connecting lines between marks are smoothed by employing the B-spline interpolation. The light-colored areas bounded by the dotted lines with appropriate color show the confidence intervals. The subfigures displaced at left bottom show corresponding relative parallel efficiency defined in Eq.~\eqref{etaC}. Note that the subfigures use log scales in (A) and (B), but linear scaling in (C) and (D) to for better illustration.}
	\label{fig:2}
\end{figure}

The comparison between the RBE and PPPM was carried out by using the MD engine of LAMMPS on the bulk water assuming the force field of SPC/E\cite{RN26}. The parameters of the PPPM are chosen automatically in LAMMPS for a given error level $\Delta$ of the relative force\cite{RN28}. The parameter $\alpha$ in the RBE is chosen to be the same as that in the PPPM. The simulations were conducted for two thousand steps to calculate the average CPU time per step, denoted as $T(C)$, where $C$ is the number of cores. Four systems of different sizes are used in the test with $N= 53367$, $311469$, $3000000$ and $100158744$ atoms, respectively, and the corresponding results are present in Fig.~\ref{fig:2}(A-D). As can be seen, the computational speed measured by $T(C)$ from the RBE can be an order of magnitude faster than that from the PPPM.

Moreover, the relative parallel efficiency $\eta(C)$ at a given number of cores $C$ defined by Eq.~\eqref{etaC} is used to characterize the scalability of the algorithm\cite{RN18},
\begin{equation}\label{etaC}
	\eta(C)=\frac{C_{\min }}{C} \cdot \frac{T_{\text {best }}}{T(C)},
\end{equation}
where $C_{\text{min}}$ denotes the minimal number of cores used in the calculation and $T_{\text{best}}$ is the run time of the fastest method at $C_{\text{min}}$. For example, $C_{\text{min}}=1$ for Fig.~\ref{fig:2}A and \ref{fig:2}B, but set to $10$ and $120$ for \ref{fig:2}C and \ref{fig:2}D, respectively, since one processor is too time-consuming and storage-limiting to simulate such large systems as in Figs.~\ref{fig:2}C and \ref{fig:2}D. The relative parallel efficiency illustrates that the RBE remains $95\%$ even for up to $10$ thousand cores when simulating $100$ million atoms (inset of Fig.~\ref{fig:2}D), significantly outperforming that of the PPPM which drops to $\sim20\%$ for the same system.

\subsection{Accuracy of the RBE and comparison with the PPPM/PME methods}
\begin{figure}[htbp]	
	\centering
	\includegraphics[width=0.5\textwidth]{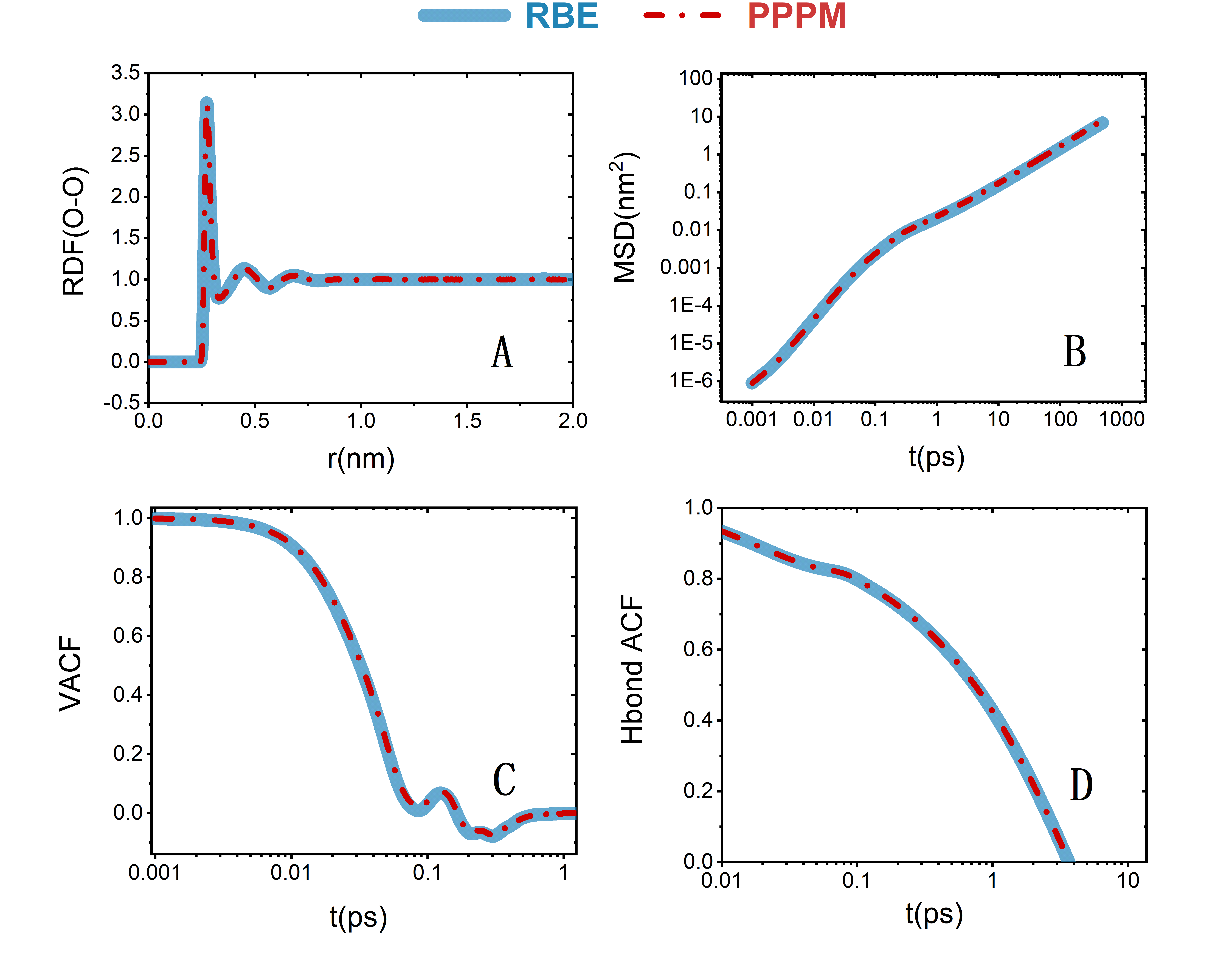}
	\caption{\textbf{Comparison of accuracy between the PPPM and RBE on bulk water}. Simulation results from the PPPM (red dash-dot line) and from the RBE with $P=100$ (blue solid line) in bulk water. (A): The radial distribution function of oxygen-oxygen in water molecules. (B): The mean square displacement of the center of mass of water molecules. (C): The velocity auto correlation function of oxygen in water molecules. (D): The hydrogen bond auto correlation function.}
	\label{fig:3}
\end{figure}

\begin{figure*}[htbp]	
	\centering
	\includegraphics[width=0.9\textwidth]{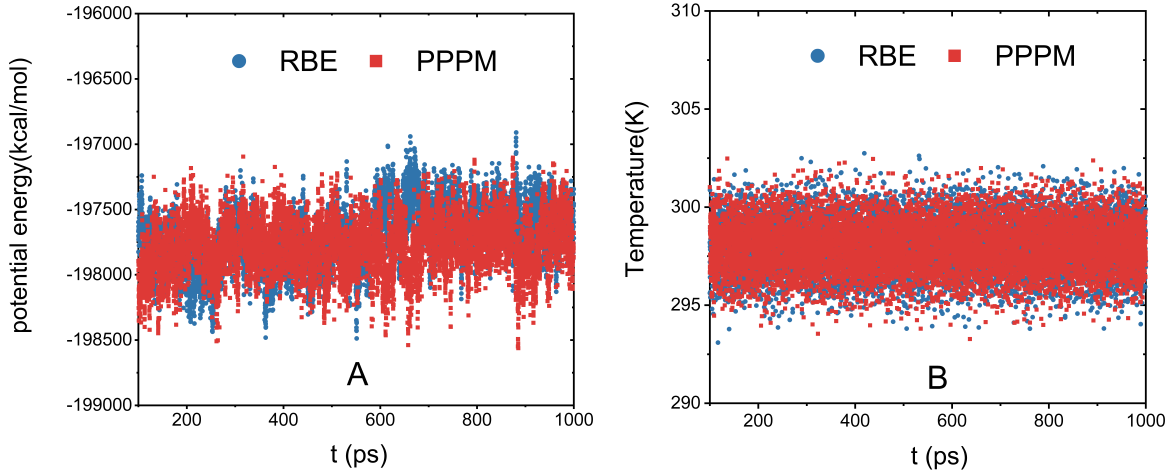}
	\caption{The variation of total potential energy (A) and the temperature variation of bulk water (B) produced by the RBE and PPPM in the NVT ensemble, respectively. As can be seen, the fluctuation of potential energy and temperature derived from the RBE and PPPM are quantitatively similar.}
	\label{fig:6}
\end{figure*}

\begin{figure}[htbp]	
	\centering
	\includegraphics[width=0.5\textwidth]{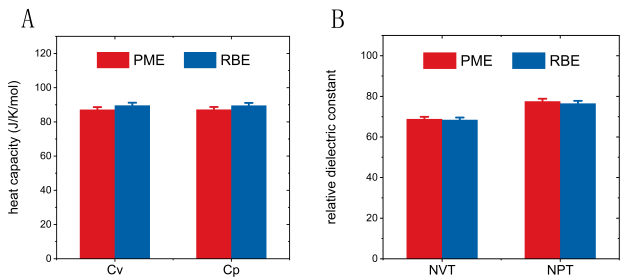}
	\caption{A: The isochoric ($C_{v}$) and isobaric ($C_{p}$) heat capacities of bulk water calculated from the trajectories of PME and RBE. B: The relative dielectric constant of bulk water from PME and RBE in the ensemble of NVT and NPT.}
	\label{fig:8}
\end{figure}

\textbf{Pure water systems}. We calculated four physical quantities on the bulk water simulation including the radial distribution function (RDF), the mean square displacement (MSD), the velocity auto correlation function (VACF), and the hydrogen bond auto correlation function (HBACF) to examine the accuracy of the RBE as compared to the PPPM. The definition of HBACF is presented in Appendix~\ref{appendix:HBACF}. The RDF of oxygen-oxygen atom pairs furnishes the spatial arrangement of water molecules. The MSD describes the translational motion on the time scale from 1 $fs$ to 1 $ns$. The VACF and HBACF characterize the short-time vibrational, liberational and rotational dynamics of water. As can be seen in Fig.~\ref{fig:3}, for both spatial arrangement ($\mathring{A}$ to $nm$) and dynamical motions of the water molecules ($fs$ to $ns$), the results derived from the RBE is almost identical to those from the PPPM. The comparisons on the fluctuation of total potential energy and the temperature of the system over simulation time are presented in Fig.~\ref{fig:6}. All the above tests were conducted at the NVT ensemble, and similar tests on the NPT were also conducted in Fig.~\ref{fig:7}. In addition, defining in Appendix \ref{appendix::capacity-dielectric}, the isochoric and isobaric heat capacities and the relative dielectric constant of bulk water in the ensemble of NVT and NPT were calculated from the RBE simulations as provided in Fig.~\ref{fig:8}, showing quantitative agreement with those derived from the PME.

\begin{figure*}[htbp]	
	\centering
	\includegraphics[width=0.9\textwidth]{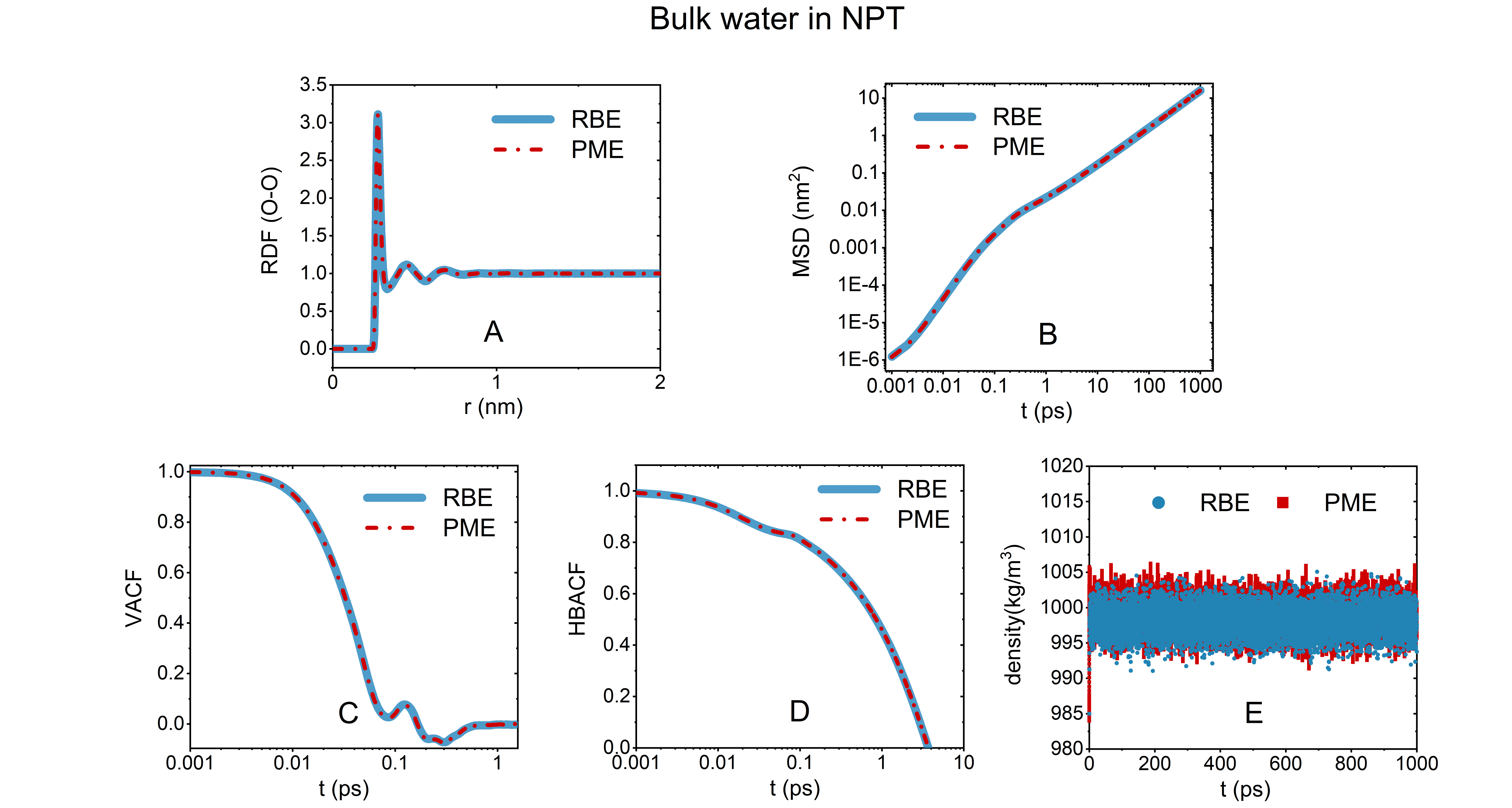}
	\caption{Comparison of accuracy between PME and RBE on bulk water in NPT ensemble. (A): RDF of O-O in H${}_2$O, (B) MSD of the center of mass of water molecules, (C): Velocity auto correlation function of oxygen atoms, (D): The hydrogen bond auto correlation, (E): Fluctuation of mass density. }
	\label{fig:7}
\end{figure*}

\begin{figure*}[htbp]	
	\centering
	\includegraphics[width=0.9\textwidth]{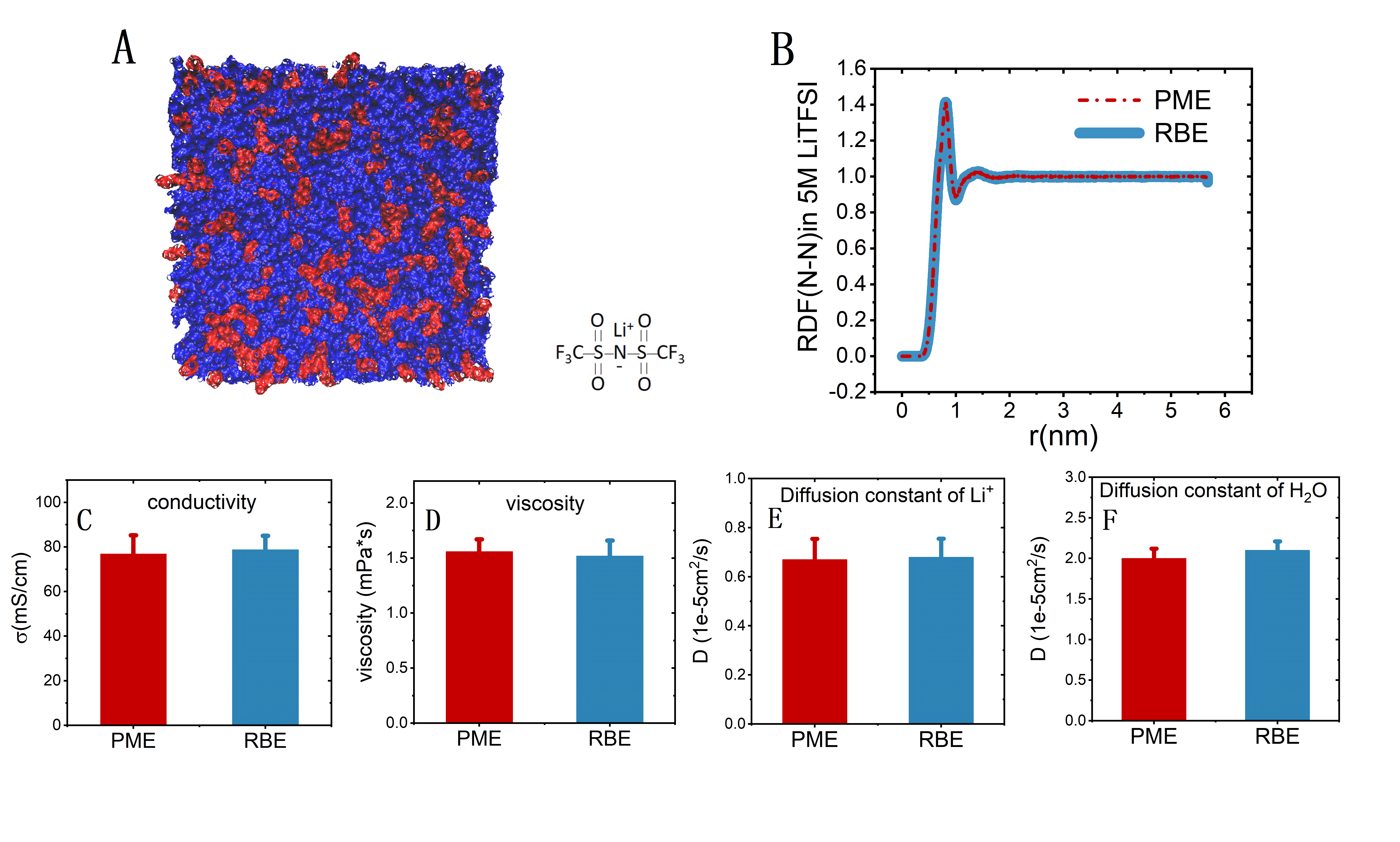}
	\caption{\textbf{Comparison of simulation results derived from PME and RBE on concentrated ionic solution}. The RBE uses $P=500$. A: Simulation snapshot of system. The anions are shown in red and water molecules are represented by blue. The inset shows the structure of the solute. B: The radial distribution function of nitrogen-nitrogen between anions. C: The conductivity of the electrolyte. D: The viscosity of the system. E: The diffusion constant of Li$^+$. F: The diffusion constant of the center of mass of water molecules.}
	\label{fig:4}
\end{figure*}

\textbf{LiTFSI ionic liquid}. The second benchmark test is an aqueous electrolyte (TFSI) at ultrahigh concentration (5 $M/L$). We implement the algorithm in GROMACS package, as the electrolyte requires a special force field which is installed in this MD engine. For comparison, the reference simulations using the PME were also conducted using GROMACS package. At such high concentration, the electrolyte is microscopically inhomogeneous, separating into two phases (water versus anions) at the length scale 1-2 $nm$, which are mutually percolated in space\cite{RN42,RN41}. Fig.~\ref{fig:4}A illustrates an MD snapshot for the electrolyte, revealing the nano heterogeneity of the system. The structural information, i.e., the RDFs of the center atom (Nitrogen) of the anions, on this concentrated electrolyte derived from both the RBE and PME are shown in Fig.~\ref{fig:4}B, while the dynamics in the system, including conductivity, viscosity and diffusion constants, are presented in Fig.~\ref{fig:4}C-\ref{fig:4}F. As can be seen, the spatiotemporal features of the system derived from the two methods are essentially the same.

\textbf{Protein in solution}. The third benchmark test is a biological sample, i.e., protein solution (see more details in the Appendix). The protein studied is lysozyme (Fig.~\ref{fig:5}A), which is a model system widely used for testing new simulation and experimental protocols for their applications on biological systems. Here, the characteristic structural information of the proteins we are testing is the root mean square deviation (RMSD) of the backbone atoms relative to the initial structure for starting the MD simulations and the surface accessible surface area (SASA) (Fig.~\ref{fig:5}B and~\ref{fig:5}C), while the dynamics of the biomolecule is characterized by the root mean squared fluctuations (RMSF) (Fig.~\ref{fig:5}D). Definitions of the RMSD, SASA and RMSF are given in Appendix \ref{appendix:SASA-RMSD-RMSF}. Moreover, we also examine the functional phase space sampled by the two methods. Lysozyme is a two-domain protein, where the two domains conduct a hinge-bending motion in order to facilitate the enzyme to break down the bacterial cell wall\cite{RN46}. The distance between two residues (C54 and C97) are often used to characterize the open-closing status of the two domains, whose distribution can be used to measure the broadness of the functional phase space and presented in Fig.~\ref{fig:5}E\cite{RN43}. As can be seen, the structure, dynamics, and functional space of the protein obtained by the PME are accurately reproduced by the RBE. More details on the data production are available at Appendix \ref{appendix:experimental}.

\begin{figure*}[htbp]	
	\centering
	\includegraphics[width=0.9\textwidth]{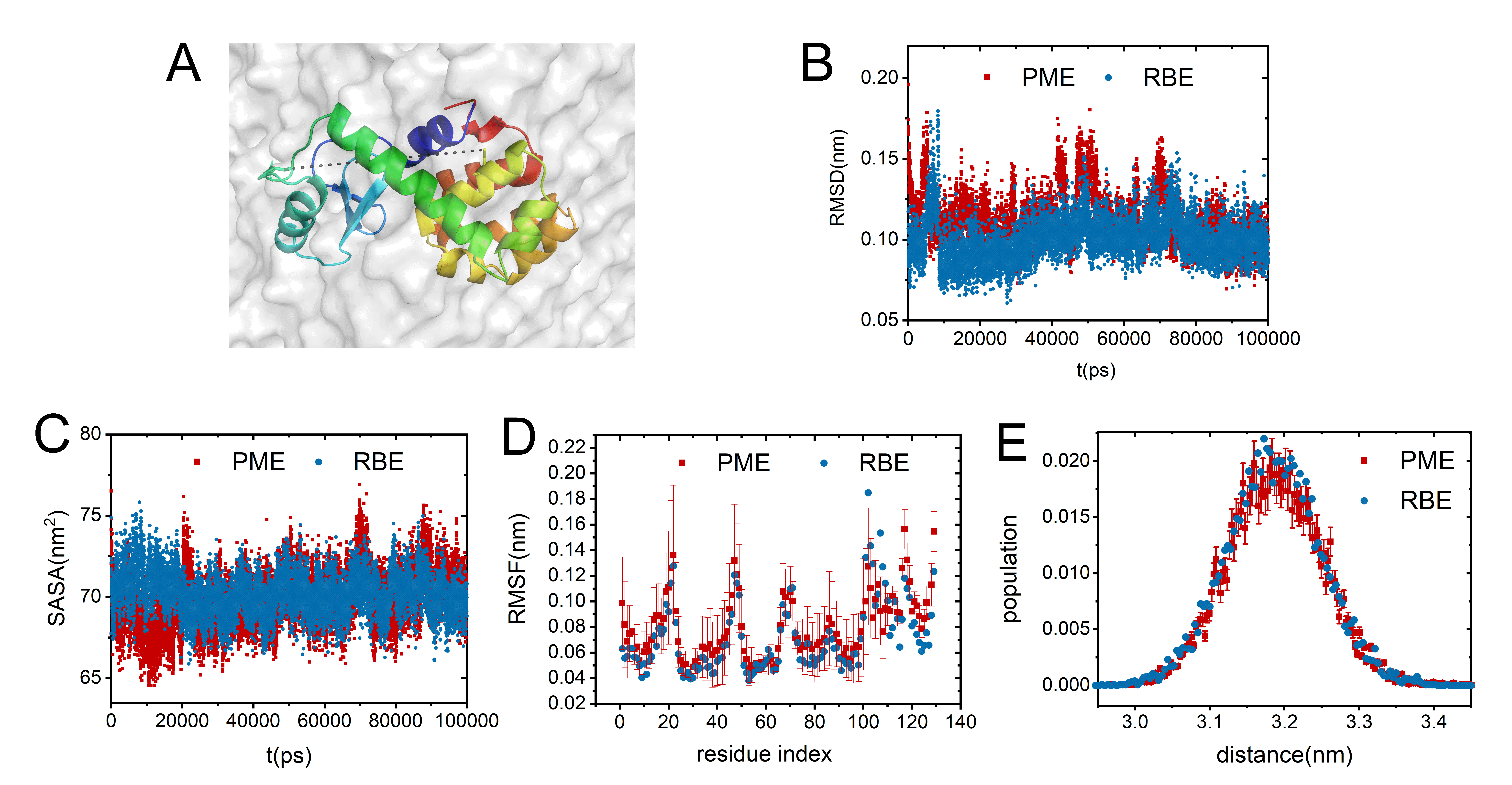}
	\caption{\textbf{Comparison of simulation results derived from PME and RBE on protein solution}. The RBE (blue) uses $P=500$. A: Simulation snapshot of the system. B. The root mean square deviation (RMSD) of the backbone atoms in protein. C. Solvent accessible surface area (SASA). D. The root mean square fluctuation (RMSF) of residues in protein. E. Distribution of the characteristic inter-domain distance between Residue C54 and C97.}
	\label{fig:5}
\end{figure*}

\section{Discussions}\label{discussions}
The RBE method inherits the advantages of Ewald-based methods but employs the random mini-batch idea and importance sampling technique to calculate the Fourier series of the long-range interaction. As a result, it achieves $O\left(N\right)$ computational complexity and gains almost linear scalability for parallel computing, outperforming existing electrostatic algorithms. When very few computational cores were employed, the acceleration of the RBE mainly comes from the relatively small batch size of $\bm{k}$ in calculating $\bm{F}_{i,1}$. With the increase of the number of CPU processors, the acceleration mainly comes from less communication. This is because six sequential rounds of communication are generally required to perform forward and backward Fourier transforms for classical FFT, whereas only one global communication is required for the RBE with $O(1)$ data transfer.

When the system is large (e.g., $100$ million atoms in Fig.~\ref{fig:2}D), the parallel efficiency of the RBE remains over $95\%$ for $10$ thousand CPU cores. When the number of atoms assigned to each processor becomes smaller (Fig.~\ref{fig:2}A-\ref{fig:2}D), the parallel efficiency of the RBE goes down but remains much better than that of the PPPM. The reduction in the RBE efficiency in small systems results from the following reasons. First, the most time-consuming part transforms gradually from intra-processor calculation to inter-processor communication for the small system. Second, the frequency of “loop remainder” in the vectorization increases while the number of particles/batches is not the multiple of the vector width. Different instructions can be used to handle the “loop remainder” in order to prevent out-of-bounds memory access along with other issues, however additional costs are introduced. Third, the cost for other parts, such as memory load and load balance, will become significant when the one consumed for arithmetic operation of the Coulomb interactions becomes small.

In addition, the speedup of the RBE is constrained by the cost of the real-space calculation, i.e., the computation of $\bm{F}_{i,2}$. In many works employing FFT-based method as their electrostatic solver, the real-space cutoff is balanced such that the costs of real and Fourier spaces are approximately the same. The real-space cutoff for the RBE should be made smaller when one accelerates the calculation in the Fourier space. The calibration of the optimized short-range algorithm for the non-covalent bonds is our ongoing project. Furthermore, if other optimized techniques for real-space cutoffs are combined with the RBE, it is possible to obtain even better acceleration. The research of all-atom simulations coupling the RBE and the recent-developed random batch list (RBL) method \cite{liang2021random} shall be studied in our subsequent works.

The time comparisons provided in Fig.~\ref{fig:2} are only for the calculation of long-range interactions. In practice, the calculation of the LJ force is often done together with the Coulomb force for saving time. Other operations, including thermostat, bond angle, construction of neighbor list, data statistics, timekeeping and diagnostic routine, have different requirements in various systems. Generally, costs of these parts are not obvious but will have considerable impact on top of the acceleration of long-range interactions. We look forward to optimize these components in mainstream packages which may become bottleneck in the future.

Moreover, the RBE method discussed here is different from any coarse-grained or enhanced sampling methods, e.g., Gaussian accelerated molecular dynamics\cite{RN51,RN63}, replica exchange\cite{RN54,RN54}, umbrella samplings\cite{RN65,RN64}, meta dynamics\cite{RN66}, as it mainly provides an efficient solution for calculating the Coulomb interactions and accelerates the simulation without losing any dynamical information on all time and length scales. It thus can be well combined with any coarse-grained and enhanced sampling methods to further speed up and scale up the simulation systems.

It is remarked that our exploration of the RBE method is limited to full periodic boundary condition. If the system is partially periodic in some directions with Dirichlet or dielectric interface conditions in other directions (e.g., the slab geometries), we believe the extension of our method is straightforward by introducing techniques developed for such problems (see Ref.\cite{liang2020harmonic,maxian2021fast,yuan2021particle} and references therein).

In summary, we have reported an efficient RBE algorithm to evaluate the Coulomb interactions in all-atom molecular dynamics simulations, and demonstrated that it can greatly improve the computational efficiency and scalability for large-scale simulations in supercomputer while maintaining the same level of accuracy. These advantages of the RBE algorithm owe to the introduction of the random mini-batch idea, which avoids the use of the FFT and significantly reduces the communication cost and the computational complexity in parallel computation. This novel algorithm will be promising for MD simulations in modern architecture and communication protocol.

\section*{Acknowledgements}
The authors acknowledge the Center for High Performance Computing at Shanghai Jiao Tong University for the computing resources. All the authors are supported by the Shanghai Science and Technology Commission (20JC1414100). J. L., Y. Z., and Z. X. are supported by the National Natural Science Foundation of China (12071288). P. T. and L. H. are supported by the National Natural Science Foundation of China (grants 11974239 and 31630002), the Innovation Program of
Shanghai Municipal Education Commission, and the Shanghai Jiao Tong University
Multidisciplinary Research Fund of Medicine and Engineering (YG 2016QN13). S. J. is supported by the National Natural Science Foundation of China (12031013). L. L. is supported by the National Natural Science Foundation of China (grants 11901389 and 11971314) and the Shanghai Sailing Program (19YF1421300).

\section*{Competing Interest Statement}
All authors have declared no competing interest.

\section*{Data Availability}
The data that support the findings of this study are available from the corresponding author upon reasonable request.

\appendix

\section*{Appendix}
\section{Experimental design}\label{appendix:experimental}
The bulk water system employs the classical SPC/E model in LAMMPS (version 7Aug2019). The SPC/E specifies a 3-site rigid water molecule with charges and Lennard-Jones (LJ) parameters assigned to each of the 3 atoms. Electrostatic interaction is modeled using Coulomb's law, and the dispersion and repulsion forces use the LJ potential. Four cubic simulation boxes of different sizes: 8.16 $nm$ (53367 atoms), 14.61 $nm$ (311469 atoms), 31.09 $nm$ (3000000 atoms), and 101.01 $nm$ (100158744 atoms) respectively were used and specified with periodic boundary conditions. The equilibration process was carried out for 500 $ns$ in NPT ensemble at 298 $K$ and 1 $bar$ with the PPPM, followed by 200 $ns$ NVT production MD simulation for data collection with the PPPM and RBE, respectively. The time integration is performed on Nosé-Hoover style non-Hamiltonian equations of motions at a temperature coupling time parameter $T_{\text{damp}}=5$ $fs$, and the scheme in LAMMPS closely follows the time-reversible measure-preserving Verlet and rRESPA integrators derived in Ref.\cite{RN29}. The velocity is initially generated according to a Maxwell distribution function at 298 $K$. All chemical bonds are converted to constraints using the SHAKE algorithm to allow a time step of 1 $fs$\cite{RN44}. During the equilibration process, the short-range part of the Coulomb interaction and the LJ interaction each with a cut-off parameter of 0.9 $nm$ are considered with periodic boundary conditions. The splitting parameter $\alpha$ of the RBE is the same as PPPM’s automatic tuning value, and the number of mini-batch is set to $P=100$. Some important physical properties are investigated to compare the RBE with the PPPM. The RDF, describing how the density of surrounding matter varies as a function of the distance from a point, is a frequently-used measurement to analyze the structure of the system\cite{RN31}. The simulation test in NPT ensemble with RBE was carried out at 298 $K$ and 1 $bar$, using Nosé-Hoover thermostat and C-rescale barostat with the coupling time 0.1 $ps$ for temperature coupling and 1 $ps$ for pressure coupling, in GROMACS (version 2021.1). The cutoff radius of the short-range Coulomb interaction and Lennard-Jones is 1.2 $nm$ with splitting parameter $\alpha=4.2$. The number of mini-batch is set to $P=100$.

The LiTFSI ionic liquid employs the OPLS-AA\cite{RN15} force field for Li${}^+$, the TIP3P model for water molecules\cite{RN45}, and the force field \cite{RN30} developed for TFSI${}^-$. The system is equilibrated in the NPT ensemble with the PME at 298 $K$ and 1 $bar$ for 500 $ns$, followed by 200 $ns$ production MD in the NVT using Nosé-Hoover thermostat with the PME and RBE, respectively. The system contains 126424 atoms, including 2560 Li${}^+$, 2560 TFSI${}^-$ and 28488 H${}_2$O. A cubic simulation box of size 11.46 $nm$ was initially used with periodic boundary conditions in GROMACS (version 2020.4). The cutoff radius of the short-range Coulomb interaction and Lennard-Jones is 1.2 $nm$ with splitting parameter $\alpha=4.2$. The number of mini-batch is set to $P=500$.

The protein solution employs charmm27\cite{RN15} force field for Lysozyme molecules, and the TIP3P model for water molecules\cite{RN45}. The system contains 38376 atoms, including 12136 water molecules and the rest protein molecules. Additional 0.1 $M$ NaCl is added into the system to describe the physiological condition. The system is equilibrated in the NPT ensemble with the PME at 298 $K$ and 1 $bar$ for 500 $ns$, followed by 200 $ns$ production MD in the NVT using Nosé-Hoover thermostat with the PME and RBE, respectively. A cubic simulation box of size 7.3 $nm$ was initially used with periodic boundary conditions in GROMACS (version 2020.4). The cutoff of the short-range Coulomb interaction and Lennard-Jones is 1.2 $nm$ with splitting parameter $\alpha=4.2$. The number of mini-batch is set to $P=500$.

\section{Hardware and Software}\label{appendix:hardware}
The computations in this paper were run on the $\pi$ 2.0 cluster supported by the Center for High Performance Computing at Shanghai Jiao Tong University. Each CPU node contains two Intel Xeon Scalable Cascade Lake 6248 (2.5GHz, 20 cores) and 12 $\times$ Samsung 16GB DDR4 ECC REG 2666 memory. The tests using 10000 CPU cores in this paper employ 250 such nodes. The computing networks are connected using 100Gbps Intel Omni-Path which is a high-speed interconnection network technology, and with this network the communication cost of both the RBE and PPPM are significantly reduced. We believe that the hardware employed in such a way accurately reflects the proportion of the cost of near-field and far-field on modern computer cluster architecture. The intel-parallel-studio/cluster.2020.1-intel-19.1.1 is used as the compiler and the LAMMPS is compiled using ``make intel$\_$cpu$\_$intelmpi''. The GROMACS is compiled using the same Intel package.
\section{The definition of the hydrogen bond auto correlation function (HBACF)}\label{appendix:HBACF}
We define the HBACF by the following function\cite{RN50}
\begin{equation}
	HBACF(t)=\frac{\langle h(0) h(t)\rangle}{\langle h\rangle} d t
\end{equation}
where the variable $h(t)$ is one, if a specified pair of water molecules is hydrogen bonded at time of $t$, otherwise it is zero. The hydrogen bond is determined by the geometric criterion, i.e., when the distance between the donor hydrogen of one water molecule and the acceptor oxygen of another water is smaller than $0.35$ $nm$, and the angle of hydrogen-donor-acceptor is smaller than $30$ degrees. The bracket means averaging over time and all the pairs of water molecules.

\section{The calculation method of heat capacity and dielectric constant}\label{appendix::capacity-dielectric}
The isochoric heat capacity ($C_v$) is calculated by the equation
\begin{equation}
		C_{v}=\frac{\sigma_{E}^{2}}{2 k_{\text{B}} T},
\end{equation}
where $\sigma_E$ is the standard deviation of total energy in the system in the NVT ensemble, $k_{\text{B}}$ is Boltzmann constant, and $T$ is the temperature. The isobaric heat capacity ($C_p$) is calculated by the equation
\begin{equation}
	C_{p}=\frac{\sigma_{H}^{2}}{2 k_{\text{B}} T},
\end{equation}
where $\sigma_H$ is the standard deviation of the enthalpy in the system in the NPT ensemble. The dielectric constant is calculated by the equation
\begin{equation}
	\varepsilon_{r}=1+\frac{4 \pi}{3 k_{\text{B}} T V}\left(\left\langle M^{2}\right\rangle-\langle M\rangle^{2}\right),
\end{equation}
where $V$ is the mean volume of the system and $M$ is the total dipole moment of the system.

\section{The calculation method of SASA, RMSD and RMSF of protein}\label{appendix:SASA-RMSD-RMSF}
The solvent accessible surface area (SASA) of protein molecule is calculated by the method of Ref.\cite{RN49}, rolling a sphere with a radius of solvent probe $1.4\mathring{A}$ over the surface of protein.

The root mean square deviation (RMSD) of a certain structure ($t$) to a reference structure is calculated by least-square fitting the structure to the reference structure, and subsequently calculating the RMSD as
\begin{equation}
	R M S D(t)=\left[\frac{1}{M} \sum_{i=1}^{N} m_{i}\left|r_{i}(t)-r_{i}(0)\right|^{2}\right]^{\frac{1}{2}}
\end{equation}
where $M=\sum_{i=1}^{N}m_i$ and $r_i\left(t\right)$ is the position of atom $i$ at time $t$.

The root mean square fluctuation (RMSF) is the standard deviation of atomic positions after least-square fitting to a reference structure,
\begin{equation}
	R M S F(i)=\sqrt{\frac{1}{\mathcal{T}} \sum_{t=1}^{\mathcal{T}}\left(r_{i, t}-\left\langle r_{i}\right\rangle\right)^{2}}
\end{equation}
where $r_{i,t}$ is the position of atom $i$ at time $t$, $\langle r_i\rangle$ is the average position of atom $i$, and $\mathcal{T}$ is the total simulation time of the trajectory.


\end{document}